\documentclass{article}

\PassOptionsToPackage{numbers}{natbib}  

\usepackage[preprint]{neurips_2024_MAIN}

\usepackage[utf8]{inputenc} 
\usepackage[T1]{fontenc}    
\usepackage{hyperref}       
\usepackage{url}            
\usepackage{booktabs}       
\usepackage{amsfonts}       
\usepackage{nicefrac}       
\usepackage{microtype}      
\usepackage{xcolor}         

\usepackage{graphicx} 

\usepackage{array}
\usepackage{tablefootnote}

\newcolumntype{P}[1]{>{\raggedright\arraybackslash}p{#1}}

\title{{The Role of Governments in Increasing Interconnected Post-Deployment Monitoring of AI}}

\author{
  Merlin Stein \\
  University of Oxford \\
  Oxford, UK \\
  \texttt{merlin.stein@bsg.ox.ac.uk} \\
  \And
  Jamie Bernardi \\
  London, UK\\
  \texttt{contact@jamiebernardi.com} \\
  \And
  Connor Dunlop \\
  Ada Lovelace Institute \\
  London, UK \\
  \texttt{cdunlop@adalovelaceinstitute.org} \\
}

\begin{document}

\maketitle

\begin{abstract}
Language-based AI systems are diffusing into society, bringing positive and negative impacts. Mitigating negative impacts depends on accurate impact assessments, drawn from an empirical evidence base that makes causal connections between AI usage and impacts.
Interconnected post-deployment monitoring combines information about model integration and use, application use, and incidents and impacts. For example, inference time monitoring of chain-of-thought reasoning can be combined with long-term monitoring of sectoral AI diffusion, impacts and incidents.
Drawing on information sharing mechanisms in other industries, we highlight example data sources and specific data points that governments could collect to inform AI risk management.

\end{abstract}


\section{Interconnected Post-Deployment Monitoring of AI as a Government Priority}
\label{sec:intro}

\textbf{People are increasingly exposed to AI systems in all areas of life. }Language-based AI systems are general-purpose technologies~\cite{eloundou2023gptsgptsearlylook}, meaning they may be deployed across contexts. Systems like GPT-4, Claude, and Gemini are increasingly being integrated into workflows at Fortune 500 companies~\cite{a16zUsage2024}, public services~\cite{ukPublicServices}, and in critical sectors like courts~\cite{Pietropaoli2023, LawGazette2024} and health services~\cite{LeeMedical2023}.

\textbf{Governments and the public have limited visibility into AI systems use and impacts.} While many applications are beneficial, adopting language-based AI systems also carries societal risks~\cite{weidinger2021ethicalsocialrisksharm, ChanHarms2023, gabriel2024ethicsadvancedaiassistants}. Applicants may be discriminated against based on their names, as recruiters screen CVs with AI systems \cite{bloombergCV2024}; certain people's jobs may be displaced~\cite{mckinseyJobs2023, usCensusSurvey2024}, and citizens' data can be more readily stolen through AI-assisted cyber attacks~\cite{fang2024teamsllmagentsexploit}. Despite these risks, very little information about how AI is used and its impacts on society is available to governments or the general public~\cite{researcherAccess2024}, which could allow harms to propagate unaddressed.

\textbf{Pre-deployment information is insufficient.} To understand risks arising from AI systems, governments and civil society have primarily developed mechanisms for gathering pre-deployment information, such as model evaluations~\cite{shevlane2023modelevaluationextremerisks}. However, pre-deployment information can not fully predict the downstream impacts of AI systems~\cite{weidinger2023sociotechnicalsafetyevaluationgenerative}. Risks ultimately arise from real-world usage, and depend on complex interactions of AI systems with people and society. For instance, combining systems with other tools can expand AI systems' capabilities in unpredictable ways~\cite{davidson2023aicapabilitiessignificantlyimproved}.

\textbf{Interconnected post-deployment monitoring can improve AI risk management by using data to inform mitigations.} By monitoring AI's actual usage and impact, researchers can derive risk taxonomies~\cite{WeidingerTaxonomy2022, koessler2023riskassessmentagicompanies} and acceptable risk tolerances ~\cite{koessler2024riskthresholdsfrontierai}. These inform the prioritisation of AI risk mitigations~\cite{khlaaf2023comprehensive, bernardi2024societaladaptationadvancedai}. \textit{Interconnected post-deployment monitoring} means 1) linking different kinds of post-deployment information for more accurate risk assessments, and 2) linking post-deployment information to specific risk mitigations. For example, linking incident data to usage data - like OpenAI's o1 chain-of-thought reasoning logs~\cite{openaio1} - could inform appropriate model safeguards or deployment corrections~\cite{oBrienEeWilliams2023} to prevent e.g., large-scale misinformation and persuasion.

\textbf{Post-deployment monitoring has been at least partly effective in other industries, and more effective when integrated into follow-up processes.}  The US Food and Drug Administration monitors population-level impacts of drugs linked to individual doctor observations~\cite{stein2023safe}; this helps it to apply new warning labels or, in the extreme case, remove a product from the market. Incident reporting in healthcare works best when connected with standardised corrective actions~\cite{stavropoulou2015effective, goekcimen2023addressing}. Accident monitoring and investigations by transport safety boards has sharply reduced fatalities across modes of transport, but only in high-income countries~\cite{fielding2011ntsb}. The EU’s Digital Services Act 2022 monitors content moderation decisions and aims to link them to structural levels of misinformation~\cite{nenadic2023structural, eu2022digitalservices}.

\textbf{Current post-deployment monitoring of AI systems is driven by civil society, with limited capacity.} Civil society organisations and researchers have revealed incidents, misuses and adverse impacts of AI systems~\cite{McGregor_2021, groves2024auditing}. While civil society plays an important role, restricted access to industrial information usually poses limits on its ability to audit industry~\cite{raji2022outsider}. AI companies partly screen usage data and customers~\cite{openai2024disruptingdeceptive, openai2024disruptingmalicious}, though incentives remain limited for publicly sharing information and tools that assist with post-deployment monitoring~\cite{euractiv2024openai, nytimes2024openaihack}. Given AI companies' limitations, governments appear to best placed to take a lead role in ensuring interconnected post-deployment monitoring. 

Consequently, this position paper argues that governments need to take an active role in conducting and incentivising post-deployment monitoring. Specifically, governments play a particular role in ensuring interconnected monitoring through facilitating information sharing and linking it to risk management. We contribute an overview of post-deployment monitoring (Section~\ref{sec:post-deploy}), a description of its challenges (Section~\ref{sec:challenges}) and recommendations for governments including specific data points to request based on successes in other industries (Section~\ref{sec:rec}).\footnote{These practices can be implemented in every  jurisdiction that regulates AI systems. However, we draw on examples in the EU, the US and the UK throughout this article. We focus on general-purpose, language-based AI systems, while most recommendations are applicable to other AI systems too.}

\section{What is Post-Deployment Monitoring of AI Systems?}
\label{sec:post-deploy}

\begin{figure}
  \centering
  \includegraphics[width=1.000\linewidth]{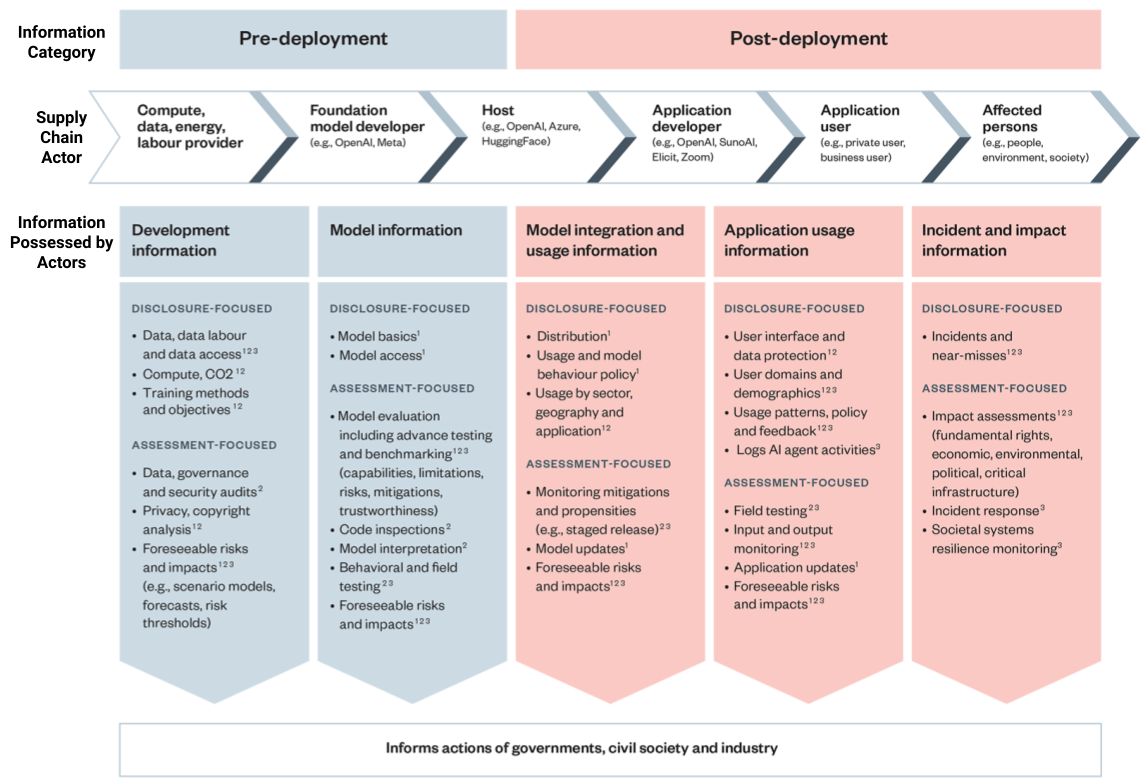}
  \caption{Information types for AI governance, categorised by supply chain actors. Some information sharing involves structured documentation (disclosure-focused), some requires additional analysis (assessment-focused). From Stein and Dunlop~\cite{steinDunlopAda2024}. Information subcategories are superscript, and are drawn from 1) the Foundation Model Transparency Index~\cite{bommasani2024foundation}, 2) the International Scientific report on the Safety of Advanced AI~\cite{internationalTechnicalReport} and 3) the Sociotechnical Safety Evaluation Repository~\cite{weidinger2023sociotechnicalsafetyevaluationgenerative}.}
  \label{fig:post-monitoring}
\end{figure}

Post-deployment monitoring increases visibility into AI models' integration into applications, usage of AI applications, and AI applications' impacts on people and society. In Figure~\ref{fig:post-monitoring}, we categorise post-deployment information by supply chain actors.

\subsection{Types of Post-Deployment Information}
\label{id:h.b0nagbn8psmj}

\textbf{Model Integration and Usage Information} 
relates to how AI models are integrated into digital applications. It includes information about how AI models are made available on the market, which application providers use them, and which industries most readily adopt AI models and downstream applications. An example is the US Census Bureau's survey of businesses' AI use~\cite{usCensusSurvey2024}.

Governments and the public have little visibility into how different sectors deploy AI systems. This hinders the ability to monitor cross-cutting risks like over-reliance on AI in certain sectors or geographies, or market concentration and unequal access. Integration information can indicate when and where these cross-cutting risks might emerge.

\textbf{Application Usage Information} relates to how an application is used in-context. It is generated when users interact with applications, ideally in the real world. It includes, for example, analysing AI system logs \cite{zhao2024wildchat}, monitoring feedback about AI applications (e.g. model vulnerabilities~\cite{ukGovtModelVulnerabilityReports, usGovtModelVulnerabilityReports, openAIVulnerability}), or conducting explicit sociotechnical field tests \cite{Schwartza2024ARIA, weidinger2023sociotechnicalsafetyevaluationgenerative}. Application usage data could also be gathered by monitoring online content for the appearance of AI outputs. Gathering usage data would be aided by requiring AI watermarks~\cite{Srinivasan2024}, content provenance~\cite{c2pa2021} and AI agent activity logs~\cite{naihin2023testing, chanAgentVisibility2024} (Section~\ref{sec:rec:4}).

Usage information is especially useful in industries requiring high levels of reliability, safety and assured benefits. Understanding how AI systems are used in real-world contexts is an important link in the causal chain towards intervening on AI's impacts.~\cite{weidinger2023sociotechnicalsafetyevaluationgenerative}. Usage monitoring could find, for example, that a few AI systems are used extensively in CV screening across companies, which might correlate discrimination risks~\cite{bloombergCV2024}. It may also show an over-reliance on AI systems for specific tasks, e.g., in critical infrastructure, which could then be reduced to prevent incidents.

\textbf{Impact and Incident Information} concerns tracking AI applications' societal effects, and adverse events and near-misses. It might be obtained through incident monitoring and reporting (see Section~\ref{sec:rec:1}), survey of affected populations~\cite{chaturvedi2023social, de2024ai}, observing socioeconomic indicators such as income disparity or employment rates~\cite{eloundou2023gptsgptsearlylook}, or monitoring societal systems and infrastructure.

\subsection{Deployment configurations: Different Supply Chain Actors’ Possession of Information}
\label{sec:post-deploy:configs}

\begin{figure}[h]
  \centering
  \includegraphics[width=1.000\linewidth]{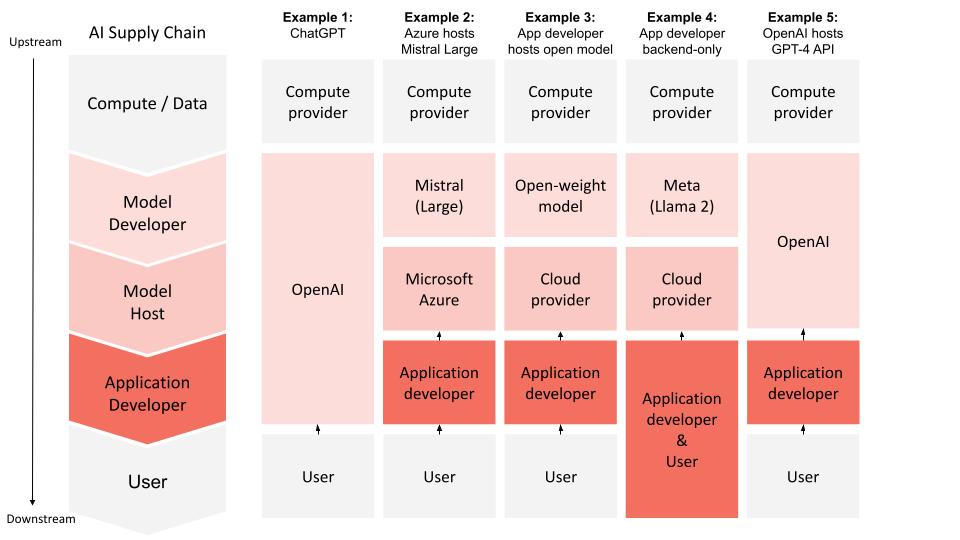}
  \caption{Example deployment configurations including some notable AI systems. On the left, we present the foundation model supply chain from~\cite{JonesExplainer2023}. On the right are five example deployment configurations (non-exhaustive).}
  \label{fig:post-deploy-examples}
\end{figure}

A single entity can fulfil one or many roles in the supply chain. For instance, OpenAI is the foundation model developer, a host \textit{and} application provider for ChatGPT. Commercial relationships between entities affect information availability due to customers' expectation of confidentiality with their vendor. We provide non-exhaustive examples of deployment configurations in Figure~\ref{fig:post-deploy-examples}.

Model developers produce AI systems. These are made available to customers through model hosts. For instance, Mistral developed Mistral Large, which is hosted by Microsoft on its Azure servers~\cite{microsoft2024mistral}, by which an API is made available to application developers. OpenAI develops and hosts its own models (though it may rent servers from a cloud provider, behind the scenes).

AI application developers create the interfaces through which people use foundation models. For example, Duolingo Max is an application through which users interact with GPT-4~\cite{duolingo2023max}. Application developers receive users' usage data. The Model host receives and processes this data too, however users and application developers often expect privacy from model hosts. However hosts might still anonymously process usage data for safety monitoring and commercial reasons.

Information may be harder to gather when foundation model developers make their model weights openly available (e.g. Mistral Large, Meta’s Llama 2, GPT-J). Open-weight models can be hosted by cloud providers that deal with developers of open-weight models (Example 2), and by application developers by renting servers from cloud providers like Amazon Web Services or Google Cloud (Example 3). In these cases, there is still a model host that receives and processes usage data. However, in theory, application developers and individual citizens could also host open-weight models on privately owned hardware. In those cases, hosts would be diffuse, small actors, thus it would be difficult to collect integration and usage data at scale.

\section{Challenges for Governments conducting Post-Deployment Monitoring of AI}
\label{sec:challenges}

Implementing policies for post-deployment monitoring of AI systems poses challenges, some seen in other industries, and others specific to AI technologies:
\begin{itemize}
  \item \textbf{User privacy. } Users expect their AI system usage to be private, thus potentially input personal data in prompts. To monitor usage data directly, it's necessary to employ consent-based data donation~\cite{zhao2024wildchat} or privacy-preserving anonymisation and data analysis techniques~\cite{Bluemke2023ExploringTR}.

    \item \textbf{Costs and independence. }Who pays the cost of compliance with post-deployment monitoring? Industry-funded monitoring, without appropriate incentive structures, can be low quality~\cite{Spelsbergj337}. Independent third-parties require appropriate access and funding~\cite{raji2022outsider}.

    \item \textbf{Information misuse. }Collecting information about incidents and misuse could strategically inform malign actors, requiring coordinated sharing mechanisms~\cite{oBrienEeWilliams2023, kolt2024responsiblereportingfrontierai}.

  \item \textbf{Commercial sensitivity. }Information detailing the rate and distribution of AI integration may reveal opportunities for competitors. Whilst current market players keep this information private by default, limited public availability may promote wealth-creating competition~\cite{teeceDynamic1997}. Where governments have offered full confidentiality for post-market monitoring, conflicts of interest can emerge between commercial activity and public safety~\cite{lexchin2004transparency, korkea-aho2017who}.

\end{itemize}

\section{Recommendations for Governments on Post-Deployment Monitoring}
\label{sec:rec}

Post-deployment monitoring and follow-up do not happen by default. We outline four recommendations for governments and AI Safety Institutes developing post-deployment monitoring policies.

\subsection{Prioritise Incident Monitoring and Reporting with causal links to AI system use}
\label{sec:rec:1}

Incident reporting and monitoring are commonly practiced in many regulated industries~\cite{fdaDatabase, faaDatabase}, and have proven at least partially effective in managing risks~\cite{stavropoulou2015effective, fielding2011ntsb}. These practices have inspired efforts to evaluate how incident reporting could support AI risk management~\cite{csetIncidents2024, stein2023safe, adaRegulatingAI2024, cltrIncidents2024, croxton2024aiincident}. Several AI incident databases have already emerged from civil society~\cite{McGregor_2021, oecdAIM, aiaaic2024, Walker_Schiff_Schiff_2024}, collecting their data from public channels. These have already informed analyses and taxonomies~\cite{marchal2024generativeaimisusetaxonomy, bogucka2024atlasaiincidentsmobile, weiKyrie2022incidents}, and have proven to help their users quantify AI harms~\cite{FefferAIID2024}.

To be effective, AI incident reporting and monitoring processes should be designed with clear policy goals, typically one of learning or accountability~\cite{weiHeim2024Forthcoming}. These goals drive post-reporting actions, such as sharing learning to relevant stakeholders~\cite{kolt2024responsiblereportingfrontierai, cweMitre} or implementing safety measures~\cite{oBrienEeWilliams2023}. Governments are often well-suited to facilitate these processes: they have the authority to mandate reporting, act as neutral parties to encourage voluntary reporting, and can provide the resources and authority for follow-up actions.

Since it's difficult to evaluate the most effective incident reporting processes in advance, governments could adopt an iterative approach to their implementation. This would allow them to build expertise and gain insight into reporting gaps over time. As a low cost starting point, government functions that catalogue AI risks - like the UK's Central AI Risk Function~\cite{ukAIRegs2023} - could monitor public channels to collect empirical evidence of AI harm, thus quantifying their risk assessments. From there, governments could explore more involved proposals, such as developing an ombudsman for citizens to report AI harms~\cite{eyeonAIada2024}, mandating reporting for major AI incidents~\cite{csetIncidents2024}, and collating AI-related incidents from sector-specific regulators~\cite{shafferShaneIncidents2024}.

\subsection{Establish Mechanisms to Gather Post-Deployment Information}
\label{sec:rec:2}

In this recommendation we outline several non-exhaustive strategies that governments and AI Safety Institutes can employ to gather post-deployment information on AI systems and models. Their respective utilities depend on the regulatory and industry context, and the nature of the monitored AI system.

\textbf{Voluntary Information Provision and Cooperation.} Governments can gather information from AI companies through both informal and formal channels for voluntary cooperation. This can involve requests for specific statistics (like an application's user count - more examples in Table~\ref{tab:examples}), but could also involve companies providing regular aggregated data streams: the UK's Office for National Statistics receives aggregated data from payment service providers~\cite{onsEconomicSocialChange2023}, which could be a useful model for governments monitoring AI integration and usage statistics. The UK and US AI Safety Institutes have already established voluntary agreements with leading AI model developers to test their models before deployment~\cite{politico2024British, axios2024anthropic}, and this framework could be expanded to include post-deployment data. Voluntary cooperation strategies are lighter-touch and more flexible than making mandatory requests, but their success is dependent on goodwill relationships, which may incur a selection bias in which companies provide the most information to government~\cite{coglianese2004seeking}.

\textbf{Mandatory reporting through legislation.} Mandatory reporting requirements ensures broad compliance, which may be essential for obtaining safety critical information. Mandatory requests often require legislative backing. A useful framework to consider for AI-related information requests is the UK's Digital Economy Act 2017, which empowers its Office for National Statistics to mandate businesses to submit specific data through binding surveys~\cite{onsLegislation}. The EU AI Act already mandates certain post-market reporting, including metric reporting (Article 72) and documentation of serious incidents (Article 73)~\cite{euChapteriX}. An effective approach depends on governments having enough knowledge to request targeted information~\cite{coglianese2004seeking}.

\textbf{Third-Party Research and Independent Monitoring. }Academics and other third party institutions play an important role in collecting and analysing post-deployment data, however their data access is often limited to public sources~\cite{raji2022outsider}. Third parties have utilised alternative sources like SimilarWeb~\cite{similarWeb} and building independent datasets for AI usage~\cite{zhao2024wildchat}. Governments can support third party efforts through funding~\cite{aisiGrants}, providing researcher access to non-public data~\cite{eu2024statusreport}, and otherwise protecting and supporting third party investigations~\cite{raji2022outsider}.

\subsection{Request Initial Data Points and Build Analysis Capacity}
\label{sec:rec:3}

Table~\ref{tab:examples} provides a preliminary, non-comprehensive list of data points that governments could start requesting from companies in the AI supply chain. The suggested data points are based on information which has been useful in other, regulated industries.

A full effort to understand AI risks would use these data points in combination with other data sources, such as macroeconomic indicators and surveying affected populations. Together, causal connections can be inferred between observed societal impacts and the integration, usage and impact data outlined in this table. For example, the environmental impacts of AI could be inferred from inference volumes~\cite{CommonCrawl2024}. Predicting economic disparities between genders can be inferred from differing usage amounts~\cite{howington2024}.

Gathering and learning from information as a government is an iterative process of identifying an informative data point, requesting it from industry, analysing the provided data, then evaluating its usefulness to generate new lines of inquiry. Requesting and analysing information requires staff time, which governments could hire-in directly~\cite{stein2024public}, fund~\cite{aisiGrants}, or facilitate through incentivising a third-party ecosystem~\cite{raji2022outsider, weidinger2023sociotechnicalsafetyevaluationgenerative}. Despite access limitations, third party organisations should not be overlooked; in the past, they have advocated for monitoring functions and the enforcement of the Digital Services Act through analysing public data~\cite{vergnolle2023putting}.

\subsection{Support Technical Governance Methods that Increase Visibility}
\label{sec:rec:4}

As the prevalence of AI outputs increases, governments should continue to encourage adoption of visibility-building technologies like content provenance~\cite{c2pa2021} and watermarking~\cite{Srinivasan2024}. As language-based AI agents are developed and become more prevalent, governments should proactively support corresponding visibility standards~\cite{chanAgentVisibility2024}. This includes AI agents outputting \textit{identifiers}, informing companies and individuals about when they are interacting with agents, indicating which developer is accountable, and otherwise creating visibility that third-party researchers could analyse.

Visibility into AI agent behaviour may also involve analysing logs~\cite{naihin2023testing}. Researchers have preserved privacy by conducting test tasks, however technical solutions may enable monitoring of real agents~\cite{Bluemke2023ExploringTR}. In any case, government agencies should work with agent developers to understand agent behaviour and human-agent interaction early in this technology's development toidentify risks, inform technical processes that mitigate them, and surface ways that companies and individuals should adapt to the diffusion of AI agents~\cite{bernardi2024societaladaptationadvancedai}.

\section{Conclusion and Future Work}

In this paper, we have argued for the critical importance of interconnected post-deployment monitoring of AI systems by governments and AI Safety Institutes. We suggest causally connecting three kinds of post-deployment information: model integration and usage, application usage, and impact and incident data. We recommend that governments and AI Safety Institutes begin building this information ecosystem by:
\begin{itemize}
    \item Prioritising incident monitoring and reporting, with causal links to AI system use.
    \item Implementing mechanisms to gather post-deployment information.
    \item Requesting specific data points from AI companies and build analysis capacity.
    \item Supporting technical governance methods that increase visibility of AI systems.
\end{itemize}

We call on the technical and AI governance research communities and AI companies to support these measures, which requires future work on assessing the effectiveness of different post-deployment monitoring approaches and using privacy-preserving techniques to build more post-deployment datasets like WildChat~\cite{zhao2024wildchat} across different sectors and applications.


\begin{table}[h!]
  \caption{Examples data points for post-deployment monitoring.}
  \label{tab:examples}
  \centering
  \setlength{\lightrulewidth}{0.01em}
  \normalsize

  \begin{tabular}{ P{2.7cm}P{3.1cm}P{3.1cm}P{3.4cm} }
  
  \toprule
  
  \textbf{Data Point} & \textbf{Utility} & \textbf{Downsides} & \textbf{Analogies} \\

  \specialrule{0.1em}{1pt}{1pt}
  
  \multicolumn{4}{l}{\textbf{Integration and model usage information} (usually provided by model hosts)} \\

  \specialrule{0.1em}{1pt}{1pt}

  \textbf{Size of user-base}, including total inference volume. & Allocate research by measuring prevalence and growth in AI applications. & Data is coarse, and survey may suffice. & EU Digital Service Act regulation only covers platforms with > 45M active EU users~\cite{eu2022digitalservices}. \\
  
  \midrule
  
  \textbf{Usage by sector}, e.g. inference volumes by Standard Industrial Classification code. & Identify potential structural risks like over-reliance and market concentration in critical sectors. & Revealing market gaps across industries may be commercial sensitive. & The US Census Bureau collects usage information by survey to understand AI's impact on employment \cite{usCensusSurvey2024}. \\
  
  \midrule
  
  \textbf{Usage by location}, e.g. inference volume per region. & Monitor adoption effects, e.g. comparing economic outcomes with regional AI use. & Revealing market gaps across geographies may be commercially sensitive. & Regional differences are commonly measured to inform digital inclusion strategies ~\cite{SandersScanlon2021}. \\

  \midrule

  \textbf{Model host downtime}, e.g. minutes/month of unavailability. & Minimise economic and other harms from downtime as AI reliance grows. & Competitive markets already incentive minimal downtime (see 'service level agreements'). & The UK's Financial Conduct Authority monitors payment service providers' up-time (e.g., Visa~\cite{fca2024reporting}). \\
  
  \specialrule{0.1em}{1pt}{1pt}
  
  \multicolumn{4}{l}{\textbf{Application usage information} (usually provided by application developers)} \\
  
  \specialrule{0.1em}{1pt}{1pt}

  \textbf{Intended use case} of an AI request, e.g. CV screening, therapy, medical. & Prioritise regulatory response based on prevalence of use cases. & Revealing market gaps in use-cases may be commercially sensitive. & the US Food and Drug Administration monitors drug usage as part of broader evaluations~\cite{fda2024regulatory}.\\

  \midrule
  
  \textbf{Degree of tool use} in AI applications (e.g., web browser access). & Assess AI's potential to operate autonomously. & Specific tools used during inference may be proprietary information. & AI specific data point, discussed in \cite{chanAgentVisibility2024}. \\

  \midrule

  \textbf{Anonymised chat logs}, with user consent \cite{zhao2024wildchat}. & Support research on AI impacts like sycophancy, over-reliance and safeguard failures. & Important privacy concerns. User awareness of sharing causes sampling bias. & The UK's Office for National Statistics receives anonymised payment data from providers~\cite{onsEconomicSocialChange2023}. \\

  \specialrule{0.1em}{1pt}{1pt}

  \multicolumn{4}{l}{\textbf{Incident and impact information} (usually better informed by observation)} \\

  \specialrule{0.1em}{1pt}{1pt}

  \textbf{Misuse statistics}, e.g. declined requests and account closures. & Measure scale of misuse and safeguard efficacy. & Reporting may create incentives to under- detect misuse. Misuse info informs attackers. & EU Digital Service Act transparency on moderation decisions and incidents \cite{Nannini2024}. \\

  \midrule

  \textbf{Incident monitoring and reporting} to identify or quantify harm. & Prevent repeated AI failures by informing legislation or safeguards  \cite{McGregor_2021}. Respond to crises. & Compliance costs, and difficulty scoping an AI incident. & Incident reporting has precedent in multiple industries \cite{csetIncidents2024, cltrIncidents2024}. \\

  \bottomrule

  \end{tabular}

\end{table}

\section*{Societal Impacts Statement}

This paper aims to increase visibility of AI's societal impacts. However, increasing visibility naturally raises privacy concerns. In this paper, the most privacy-sensitive policy we discuss is the analysis of users' chat logs to help understand AI usage (other metrics we discuss are usually aggregated and/or carry no personal information, only high-level statistics about usage). Analysis of usage is already conducted by foundation model providers for misuse monitoring, and is usually highly automated (meaning few humans require access to chat logs). Monitoring usage information should be carried out using best practice developed for those purposes, with a minimal set of employees able to view personal data. When considering data sharing agreements, governments and other actors should: follow data protection laws in the relevant jurisdiction(s) at a minimum; ensure data sharing agreements are clear and transparent to users; and take every effort to conceal or remove personal data using privacy-preserving technologies.

\begin{ack}


For helpful conversations and comments on this work, we'd like to thank Rishi Bommasani, Simon Mylius, Tommy Shaffer Shane, Kevin Wei and Zoe Williams.

\end{ack}


\bibliography{neurips_2024_MAIN}

\end{document}